\documentstyle[12pt,psfig,draft,lscape,array]{nature-ejb}

\def\Mo{{\rm M_\odot}}
\def\simlt{\mathrel{\hbox{\rlap{\hbox{\lower4pt\hbox{$\sim$}}}\hbox{$<$}}}}
\def\simgt{\mathrel{\hbox{\rlap{\hbox{\lower4pt\hbox{$\sim$}}}\hbox{$>$}}}}
\def\ale{\mathrel{\hbox{\rlap{\hbox{\lower4pt\hbox{$\sim$}}}\hbox{$<$}}}}
\def\age{\mathrel{\hbox{\rlap{\hbox{\lower4pt\hbox{$\sim$}}}\hbox{$>$}}}}

\def\spose#1{\hbox to 0pt{#1\hss}}
\newcommand\lsim{\mathrel{\spose{\lower 3pt\hbox{$\mathchar"218$}}
     \raise 2.0pt\hbox{$\mathchar"13C$}}}
\newcommand\gsim{\mathrel{\spose{\lower 3pt\hbox{$\mathchar"218$}}
     \raise 2.0pt\hbox{$\mathchar"13E$}}}

\begin{document}

\title{\Large \bf Earth-mass dark-matter haloes as
the first structures in the early Universe}
\author{
   J.~Diemand$^1$\thanks{Present adress: Department of Astronomy  Astrophysics, University of California,
1156 High Street, Santa Cruz CA 95064, USA},
B.~Moore$^1$ \& J.~Stadel\affiliation[1]
     {Institute for Theoretical Physics, University of Zurich, 
Winterthurerstrasse 190, CH-8057 Z\"urich, Switzerland},   
}

\date{\today}{}
\headertitle{The first structures in the Universe}
\mainauthor{Diemand et al.}

\summary{
Before a redshift z=100, about 20 million years 
after the big bang, the universe was nearly smooth
and homogenous \cite{Peebles1982}. After this epoch tiny fluctuations 
imprinted in the matter distribution during the initial expansion
began to collapse via gravitational instability.
The properties of these fluctuations depend on 
the unknown nature of dark matter 
\cite{Hofmann2001,Berezinsky2003,Green2004}, 
which is one of the biggest
challenges in present day science
\cite{Jungman1996,Ellis2002,Bertone2004}.
Here we present supercomputer simulations of
the concordance cosmological model assuming
neutralino dark matter and find 
the first objects to form are
numerous earth mass dark matter halos about as 
large as the solar system.
They are stable against 
gravitational disruption, even within the central regions of
the Milky Way, and we expect over $10^{15}$ 
to survive within the Galactic halo with one passing through the
solar system every few thousand years.
The nearest structures
will be amongst the brightest sources for gamma-rays 
from particle-particle annihilation. 
}

\maketitle


The cosmological parameters of our universe and 
initial conditions for structure formation have 
recently been measured via a combination of
observations including the 
cosmic microwave background (CMB)\cite{Spergel2003} , 
distant supernovae \cite{Riess1998,Perlmutter1999}
and the large scale distribution of galaxies 
\cite{Tegmark2004}.
Cosmologists now face the 
outstanding problem of understanding the origin of 
structure in the universe from its strange mix of 
particles and vacuum energy \cite{Peebles1984,Tegmark1997}.

Most of the mass of the universe must be a non-baryonic 
particle \cite{Peebles1982,Turner2002}
that remains undetected in laboratory experiments.
The leading candidate for this ``dark matter''
is the neutralino, the lightest
supersymmetric particle which is predicted to solve
several key problems in the standard model for particle
physics \cite{Jungman1996}. This cold dark 
matter (CDM) candidate is not completely
collisionless. It can collide with baryons thus revealing its
presence in laboratory detectors, although the cross-section
for this interaction is extremely small. In a cubic meter 
detector containing $\sim 10^{30}$ baryon particles
only a few collisions per day are expected from the $\sim 10^{13}$
dark matter particles that flow through the experiment
as the earth moves through the galaxy.
The neutralino is its own anti-particle and
it can self-annihilate creating a shower of new particles
including gamma-rays \cite{Jungman1996}. 
The annihilation rate increases
as the density squared therefore the central regions of the 
Galaxy and its satellites will give the strongest signal
\cite{Lake1990,Bergstrom1999,Calcaneo2000,Prada2004}.
However the expected rate is very low - the flux of photons 
on earth is the same as we would receive
from a single candle placed on Pluto.
Numerous experiments utilising these effects are underway
which may detect the neutralino within the next decade
\cite{Bertone2004}.  
Furthermore, in the next few years the LHC at CERN will confirm or 
rule out the concepts of supersymmetry (SUSY)\cite{Ellis2002}.

We followed the growth and subsequent 
gravitational collapse and virialisation of the
first structures in the cold dark matter universe
with supercomputer calculations. 
The challenge is to accurately follow the evolution of the universe
on scales that are many orders of magnitude smaller than previously studied,
whilst also capturing the gravitational dynamics from large scales.
We use a multiscale technique \cite{Bertschinger2001}
in order to achieve the desired resolution 
within a small average density patch of the universe 
which is nested within a hierarchy of
larger and lower resolution grids of particles. 

The fluctuations are
imposed on the particles using 
accurate calculations of the linear theory power spectrum for
a SUSY model with a particle mass $m_\nu$ = 100 GeV. This
includes collisional damping, free streaming and the transfer of
fluctuations through the matter-radiation era of the universe
\cite{Hofmann2001,Berezinsky2003,Green2004}.
The resulting power spectrum is close to a power law $P(k)\propto k^{n}$ 
with $n=-3$ with an exponential cut-off at 0.6 comoving parsecs 
which corresponds to a mass scale of $10^{-6} M_\odot$. The cutoff 
scale depends on the neutralino mass and decoupling energy. From 
accelerator searches we know that $m_\nu >$ 37 GeV and the cosmic 
matter density
sets an upper limit at 500 GeV. The damping scale
for the allowed neutralino models differ from 
the model we used by less than a factor of three in mass 
\cite{Hofmann2001,Berezinsky2003,Green2004} and
therefore structure formation is very similar
in all SUSY-CDM scenarios. A less popular CDM candidate is the axion,
it has a much smaller damping scale of $10^{-13} M_\odot$.
For comparison we simulated the high resolution region with an
axion CDM fluctuation spectrum on the resolved scales. 
Both models produce equal halo abundances above $5 \times 10^{-6} M_\odot$,
but the axion model also forms bound structures
down to the smallest resolved scales, see Figure \ref{fig:massfn}.
In this letter we concentrate on the properties of the first structures
to form in the SUSY-CDM model. 

\begin{figure}
\centerline{\psfig{file=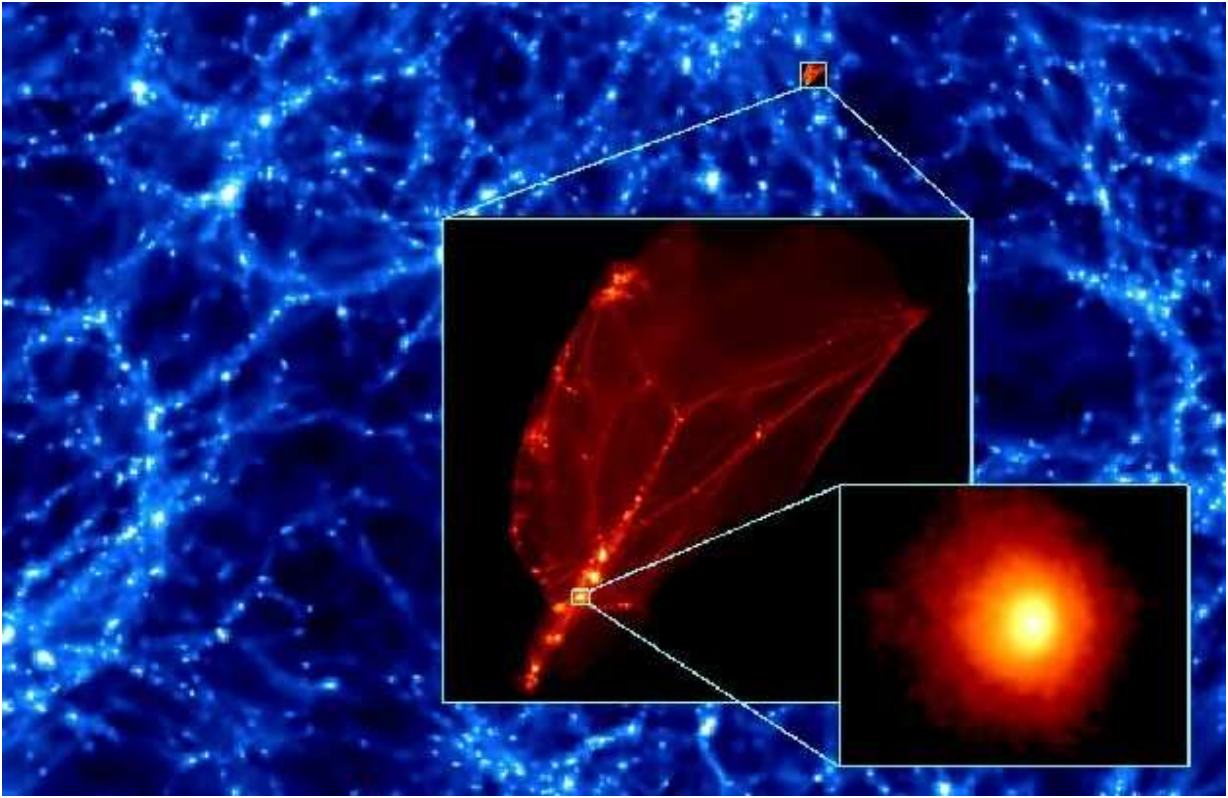,width=6.4in,angle=0}}
\caption[]{\small A zoom 
into one of the first objects to form in the universe.
The colours show the density of dark matter 
at redshift 26. Brighter colours correspond
to regions of higher concentrations of matter.
The blue background image shows the small scale structure in the top cube 
(cube size = [3 comoving kpc]$^3$) 
which has a similar filamentary topology as the 
large scale structure in the CDM universe.
The first red image zooms by a factor of one hundred into the average 
density high resolution region. This region was initially a cube
of [60 comoving pc]$^3$ resolved with 64 million particles
with a gravitational softening
of $10^{-2}$ comoving parsecs and masses
$1.2\times 10^{-10}M_\odot\equiv M_{moon}/300$.
The final image shows a close up of one of the individual dark matter halos
in this region, again zooming in by a factor of one hundred so 
that the box has a physical length of 0.024 parsecs.
This tiny triaxial Earth mass halo has a cuspy density profile and 
is smooth, devoid of the substructure
that is found within galactic and cluster mass dark matter halos.
Even though the index of the power spectrum is very steep on 
these scales, $n\approx-3$, we find that halos can collapse before
merging into a larger system, rather than the niave expectation that
all scales are collapsing simultaneously thus erasing such structures.
}
\label{fig:zoom}
\end{figure}

We evolve the initial particle distribution using a parallel multi-stepping
treecode, starting at a redshift $z=350$ when the fluctuations 
are still linear. The high resolution region forms the first 
non-linear structures at z=60 and the entire region
quickly becomes distorted by the complex tidal field from the
surrounding overdensities. By a redshift
z=26 the high-resolution region begins to merge into
the lower resolution surroundings and 
we do not analyse the region further, however this is sufficiently late that
about 5 percent of the mass in the region
has collapsed into bound dense structures (halos), 
see Figure~\ref{fig:zoom}. 

The first dark matter halos to collapse and virialise are
smooth triaxial objects of mass $10^{-6}M_\odot$ 
and half mass radii of $10^{-2}$ parsecs.
Figure 2 shows the density profiles of three representative
halos at z=26 which are well fitted
by single power law density profiles $\rho(r)\propto r^{-\gamma}$
with slopes $\gamma$ in the range from 1.5 to 2, 
similar to galactic halos shortly after their 
formation \cite{Tasitsiomi2004}. 
Note that the densities at the virial radius
are about an order of magnitude
higher than the density at $0.01 r_{\rm virial}$ in a galactic halo today, 
which makes the survival of many of these halos as galactic substructure   
possible. The central resolved densities reach $10^9$ times the
mean background density at one percent of their virial radii.
Unlike galaxy and cluster mass CDM halos, they do not contain
substructure since no smaller mass halos have collapsed in the hierarchy.

Figure 3 shows the mass function of halos. 
We use a friends of friends algorithm with a linking
length set to identify the dense central regions of collapsed
halos, then for each halo centre we recursively search for the
radius $r_{200}$ that is at an overdensity of 200 times the
cosmic mean density. The resulting 
halo mass function is steep $dn(M)/d \log M\propto M^{-1}$. 
For comparison we plot an extrapolation of 
the halo mass function found on much larger scales $> 10^{7}M_\odot$ [\cite{Reed2003a}]
which fits surprisingly well up to the cut-off scale of $10^{-6}M_\odot$
below which we find no more structures. 

At these epoch the baryons are kept sufficiently warm by the CMB that they
are unable to cool and form visible objects such as stars or planets 
within such tiny systems \cite{Tegmark1997}.
The dark halos may be detected 
via gravitational effects such as
lensing or dynamical perturbations. Although we
can not simulate the entire galactic halo at the resolution required to
determine the survival statistics of these objects we can make 
some simple estimates of their survival and abundances. 
As the Galactic halo is assembled, these first objects merge successively into
more massive systems.
From scales of $10^7M_\odot$ to $10^{15}M_\odot$
the mass function of substructure is a self similar 
power law of slope $dn(m)/dm \propto m^{-1.9}$
[\cite{Diemand2004sub}]. 
Extrapolating the subhalo mass function to the smallest scales gives us a
total number of substructure halos 
$N(M>10^{-6}M_\odot) \sim 5\times 10^{15}$ and the expected number 
density of subhalos at the solar radius is 
$n(R_\odot)\approx 500 {\rm pc}^{-3}$, assuming that they trace the mass.
Although this extrapolation is made to much smaller masses than simulated
previously, the substructure within halos 
collapsing at $z\approx 15$ with masses $\sim 10^7M_\odot$
fit the extrapolation from larger mass scales \cite{Moore2001} even though
they form from regions of the CDM power spectra with effective index 
$n\approx -2.95$.

Can these structures survive the strong disruptive gravitational forces
from the Galaxy? The tidal radius is simply the
inner Lagrange point of the rotation of the two body system. For
halos with power law density profiles $\rho(r)\propto r^{-2}$ 
we find $r_t=(R v_{sat})/(\sqrt{2} V_{parent})$
where $v_{sat}$ and $V_{parent}$ are the effective circular velocities 
($V=\sqrt{GM/r}$) of the satellite and main halo and $R$ is the pericentric distance
of the satellite. For the smallest mini-halos $v_{sat}\approx 1 {\rm m/s}$,
$r_{200}=0.01$ parsecs. 
Therefore within the Galactic potential
these halos could survive completely 
intact to about 3 kpc from the centre, well within the galacto-centric
position of the sun. 
Encounters between halos and with stars and molecular clouds may disrupt
a small fraction of these structures but using the impulsive heating
approximation we estimate that most 
will survive with little mass loss.

A significant fraction of the mass may lie within bound structures at our
location within the Galaxy, 
lowering the available smoothly distributed matter necessary
for direct detection experiments. The earth passes through a dark matter
mini-halo every 10,000 years, an encounter which lasts for about 50 years, therefore
most of the time the earth is within an underdense region of dark matter. Integrating the
mass function from $10^{-6}M_\odot$ to $10^{10}M_\odot$, normalised such that
10\% of the mass is within substructure above a mass scale of $10^7M_\odot$
(as given by simulations of Galactic halos) we find that about 50 percent of the 
mass is bound to dark matter substructures.
The velocity perturbation to a planetary orbit or satellite is very small 
($\approx 10^{-6} {\rm m/s}$), well below the observational constraints.
However resonant encounters and 
the cumulative effects of $\approx 10^6$ impulsive
encounters may cause significant perturbations to some of the 
bodies orbiting in the Oort cloud surrounding the solar system.

Compact objects in the mass range considered here could produce
a microlensing signal in a multiply lensed quasar image, such as 
time varying flux differences \cite{Schmidt1998}. 
The lensing object needs to be smaller than the Einstein radius
\begin{equation}\label{pm}
r_E = 3.7 \times 10^{16} \sqrt{\frac{M}{hM_\odot}} \; {\rm cm} \;\;\; .
\end{equation}
For a $10^{-6}M_\odot$ object $r_E \simeq 10^{-7}$ pc, which is
much smaller than the size of the mini-halos considered here, therefore
it is unlikely that gravitational lensing can provide a constraint
on their presence, either in our halo or on cosmological path lengths
to distant quasars.

Indirect detection is a more interesting possibility and several
ongoing and planned experiments aim to detect the atmospheric Cerenkov light
from gamma-rays produced by neutralino annihilation 
in the cores of dark halos
halos\cite{Bertone2004,Mori2001,Cogan2004,Hinton2004,Wittek2004}. 
Simple scaling arguments can show that minihalos can have high relative
luminosities in $\gamma$-rays.
The absolute
$\gamma$-ray luminosity of a dark matter halo with an NFW density profile
is proportional to $L\propto \rho_s^2\;r_s^3$, 
where $r_s$ is the scale radius of the NFW profile and 
$\rho_s = \rho(r_s)$ \cite{Savvas2003}. 
The relative luminosity that would arrive at the detector from 
a halo at a distance $d$ is then $L_{\rm rel} \propto L d^{-2}$.

Now we compare the relative luminosity of a minihalo at a distance of 0.1 
parsec (their expected mean separation) 
to the signal from the centre of the Draco dwarf galaxy: 
\begin{equation}
{{L_{\rm rel,mini}}\over{L_{\rm rel,draco}}}
\propto \Biggl({{7\times10^6 \rho_{\rm crit}}\over {1.7\times10^5\rho_{\rm crit}}}\Biggr)^2 
\Biggl({0.005 {\rm pc} \over 300 {\rm pc}}\Biggr)^3 
\Biggl({0.1 {\rm pc} \over 82,000 {\rm pc}}\Biggr)^{-2}  
\approx 5
\end{equation}
where we used the typical minihalo properties from our simulations. 
The large abundance of the smallest subclumps 
compensates their smaller absolute luminosity and 
the closest of them will be bright sources of $\gamma$-rays.
The background flux will be enhanced by a boost factor of over two orders
of magnitude over a smooth Galactic dark matter potential.
Current indirect detection experiments such as VERITAS\cite{Cogan2004}, 
HESS\cite{Hinton2004}, MAGIC\cite{Wittek2004} 
or CANGAROO-III\cite{Mori2001} can probe part
of the parameter space predicted by SUSY theory by observing the galactic
centre. However this region is dynamically complex since it contains 
numerous confusing astrophysical gamma ray sources and a supermassive
black hole that can erase the central cusp. CDM mini-halos are 
potentially bright and will not suffer
from these problems. All sky surveys could detect some nearby minihalos
which would have a characteristic extent on the sky that is 
similar to that expected for a more distant satellite 
galaxy like Draco.

\

{\noindent Please address all correspondence and requests for materials to Professor Ben Moore.}

\begin{acknowledge}
It is a pleasure to thank Anne Green, Dominik Schwarz, 
Philippe Jetzer, Marco Miranda, 
Andrea Maccio and Gianfranco Bertone for helpful discussions.
All computations were performed on the zBox
supercomputer at the University of Zurich.
This work was supported by the Swiss National Science Foundation.

\end{acknowledge}

\clearpage

\begin{figure}
\centerline{\psfig{file=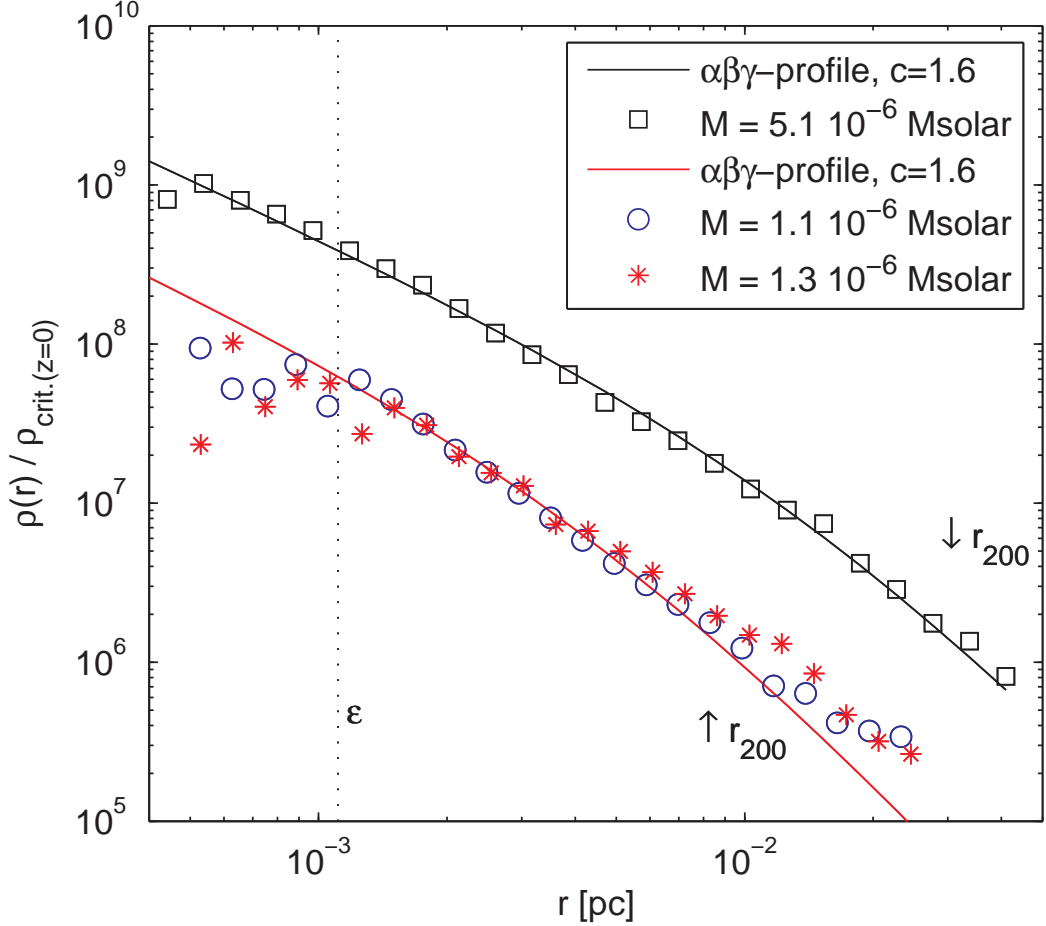,width=6in,angle=0}}
\caption[]{\small 
Radial density profiles of three typical minihalos at redshift 26. The radial
distance is plotted in physical units and we show low concentration 
$\alpha\beta\gamma$-profiles for comparison. We use the mean 
dark matter profile infered from the highest resolution galaxy cluster simulations 
\protect\cite{Diemand2004pro}, i.e. $(\alpha\beta\gamma)=(1,3,1.2)$.
The vertical dotted line indicates our force resolution and the arrows indicate
the radii that is 200 times the background density.
Across the entire range of halo masses from $10^{-6}$ to $10^1 M_\odot$,
we find small concentration parameters $c < 3$.
We do not observe a trend of concentration 
with mass, possibly because the halos all form at a similar
epoch as expected when the power spectrum is so steep. 
} 
\label{fig:pros}
\end{figure}

\begin{figure}
\centerline{\psfig{file=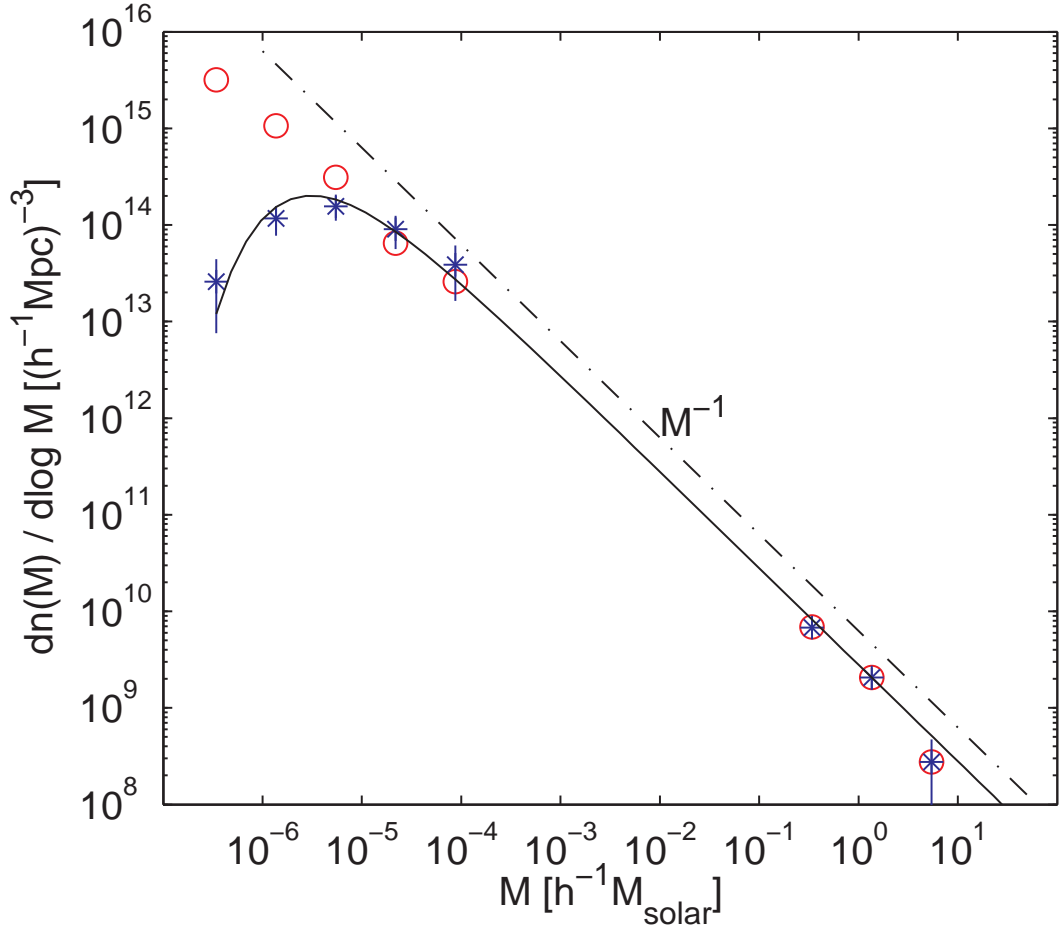,width=6in,angle=0}}
\caption[]{\small 
The abundance of collapsed and virialised dark matter halos of a given 
mass. The same region was simulated twice using different types of 
intial fluctuations: (A) SUSY-CDM with a 100 GeV neutralino
(stars) and (B) an additional model with no small scale cut-off to the power
spectrum (open circles) as might be produced 
by an axion dark matter candidate. 
Densities are given in co-moving
units, masses in $h^{-1} \Mo = 1.41 \Mo$, where $h=0.71$ is the
normalized present day Hubble expansion rate.
Model (B) has a steep
mass function down to the resolution limit whereas run (A)
has many fewer halos below 
a mass of about $5\times 10^{-6} h^{-1}\Mo = 3.5\times 10^{-6}h^{-1}\Mo$. 
(Our simulations do not
probe the mass range from about $3 \times 10^{-4} h^{-1}\Mo$ 
to $2 \times 10^{-1} h^{-1}\Mo$.)
The dashed-dotted line shows an extrapolation of the  
number density of galaxy halos (from \protect\cite{Reed2003a})
assuming $dn(M) / d \log M \propto M^{-1}$.
The solid line is the function 
$dn(M) / d \log M = 2.8\times10^9(M/h^{-1}\Mo)^{-1} 
\exp[-(M/M_{\rm cutoff})^{-2/3}] (h^{-1}Mpc)^{-3}$, 
with a cutoff mass $M_{\rm cutoff} =5.7\times10^{-6}h^{-1}\Mo$.  
The power spectrum cutoff is $P(k) \propto \exp[-(k/k_{fs}]^2)$,
where $k_{fs}$ is the free streaming scale and
assuming $k \propto M^{-1/3}$ motivates the exponent of $-2/3$ in
our fitting function.
}
\label{fig:massfn}
\end{figure}

\end{document}